\documentclass[reprint,superscriptaddress,amsmath,amssymb,aps,]{revtex4-1}
\usepackage{ulem}
\usepackage{graphicx}
\usepackage{dcolumn}
\usepackage{bm}
\usepackage[caption=false]{subfig}
\usepackage{floatrow}
\usepackage{appendix}
\usepackage{amsmath}
\usepackage[italicdiff]{physics}
\usepackage[capitalise]{cleveref}
\usepackage{xcolor}
\DeclareCaptionLabelFormat{adja-page}{\hrulefill\\#1 #2 \emph{(previous page)}}

\begin{abstract}
 We present a simplified web-based application for simulating x-ray and photoelectron spectra of transition metals, built around the notion that web-based applications lower the bar for novice users. The application provides a simple interface to simulate x-ray absorption spectroscopy, resonant inelastic x-ray scattering, and angle-resolved photoemission spectroscopy, incorporating the effects of local electronic interactions, which give rise to multiplets, spin-orbit coupling, crystal field effects, and ligand hybridization/charge transfer. Results can be obtained that highlight the key role of photon polarization.
\end{abstract}
\begin{document}
\title{Web-Based Methods for X-ray and Photoelectron Spectroscopies}
\author{Thomas P. Devereaux}
\email{tpd@stanford.edu}
\affiliation{Stanford Institute for Materials and Energy Sciences,
SLAC National Accelerator Laboratory, 2575 Sand Hill Road, Menlo Park, CA 94025, USA}
\affiliation{
Department of Materials Science and Engineering, Stanford University, Stanford, CA 94305, USA}
\affiliation{
Geballe Laboratory for Advanced Materials, Stanford University, Stanford, CA 94305, USA}

\author{Brian Moritz}
\affiliation{Stanford Institute for Materials and Energy Sciences,
SLAC National Accelerator Laboratory, 2575 Sand Hill Road, Menlo Park, CA 94025, USA}

\author{Chunjing Jia}
\affiliation{Stanford Institute for Materials and Energy Sciences,
SLAC National Accelerator Laboratory, 2575 Sand Hill Road, Menlo Park, CA 94025, USA}

\author{Joshua J. Kas}
\affiliation{University of Washington, Seattle, WA 98195-1560, USA}

\author{John. J. Rehr}
\affiliation{University of Washington, Seattle, WA 98195-1560, USA}
\affiliation{Stanford Institute for Materials and Energy Sciences,
SLAC National Accelerator Laboratory, 2575 Sand Hill Road, Menlo Park, CA 94025, USA}

\date{\today}

\maketitle

\section{Introduction}
The application of x-ray spectroscopies to investigate the nature of charge, spin, orbital, and lattice excitations in quantum materials has expanded greatly, due in part to the advances made in synchrotron and free electron laser light sources worldwide. In many cases, as energy resolution has been pushed towards the level of tens of millielectron volts or better, excitation dynamics involving states accessible within room temperature $k_BT$ and below have been revealed across a wide range of so-called ``intertwined" materials - materials in which it is insufficent to describe phenomena in terms of separate degrees of freedom. This includes   wide classes of materials relevant for superconductivity, magnetism, ferroelectricity, and energy storage to name just few. While x-ray absorption and core level spectroscopy have long been able to infer local coordination numbers and predominant charge valency configurations with atomic specificity,  advances in resonant inelastic x-ray scattering (RIXS) have opened direct access to the energy dispersion of various collective modes in quantum materials stemming from numerous orders - plasmons, magnons, orbitons, spinons, charge doublons/excitons, charge density modes, and phonons, as well as the mutual couplings among the collective modes and with underlying particle-hole excitations. Thus, x-ray spectroscopy has become the go-to tool for providing a thorough understanding of quantum material properties.

As a consequence, there has been steady growth in the number and sophistication of computer codes available to simulate, and thereby interpret the spectra  of quantum materials. Robust X-ray absorption (XAS) and emission (XES) codes have been available for a number of decades and have served as a foundation for researchers to obtain a broad understanding of the spectra. In many cases an effective one-electron model, as in the real-space Green's function FEFF codes \cite{feff10} can be a good approximation. However, for transition metal $L$-edges it is important to treat   multiplet effects stemming from both intra-valence and core-valence Coulomb interactions including crystal fields, and spin-orbit effects\cite{Cowan,Thole}. The atomic-multiplet codes {\it CTM4XAS} and {\it CTM4RIXS} have been workhorses for many researchers across chemistry, physics, and materials science\cite{CT4XAS}. The recent addition of {\it Quanty} also  incorporates dynamical mean field theory (DMFT) for lattice problems, as well as adding extensions to RIXS\cite{Quanty}, broadly expanding availability. Finally, recent codes that simulate angle-resolved photoemission spectroscopy (ARPES) are available. These codes give tutorial-level experience to users, both to familiarize and explore spectra from a simple and intuitive tight-binding perspective\cite{Chinook}. While we do not intend to review them here, a list of the many  codes currently available is given in Ref. \onlinecite{Review}.
While these codes will in no doubt continue to improve, a typical barrier  encountered by users is the overhead associated with download and set-up of many applications. Particularly in the case of researchers wishing to become familiar with techniques in a tutorial-like setting, or for students to explore spectroscopy in a class-like setting, one typically desires a simplified method that sacrifices many of the bells-and-whistles of complexity to obtain a more transparent usage.

In this communication, we introduce two web-based methods, the first for simulating XAS and RIXS ({\tt webXRS})\cite{webXRS} and the second for simulating ARPES ({\tt webARPES})\cite{webARPES} for transition metal $L$-edges, whereby users can enter a minimal set of input parameters through a web-interface and run codes hosted on internal servers, significantly lower the barriers for novice or casual users.  A related tool ({\tt webXAS}) for simulating $K-$edge XAS with FEFF10 will be discussed elsewhere\cite{feff10webapp}.

\section{Background and Formalism}
Many of the methods and background information for for $L$-edge XAS, RIXS, and ARPES are available in Refs.\ \onlinecite{Cowan,Thole,CT4XAS,Quanty,Chinook}  and \onlinecite{KotanideGroot,MvV}, so here, we only   summarize the formulae relied upon in  {\tt webARPES} and {\tt webXRS} codes.

The electron removal spectra $I(\omega)$ for incident light of energy $\hbar\omega_i$ and polarization ${\bf e_i}$, $I_{\hbar\omega_i,{\bf e_i}}({\bf k_i},\omega)$, measured in XPS/ARPES is given by
\begin{eqnarray}
I_{\hbar\omega_i,{\bf e_i}}({\bf k_i},\omega)&=&\frac{1}{Z} \sum_{i,\nu} e^{-\beta E_i} \mid\langle \nu\mid \hat D_{\bf k_i} ({\bf e_i})\mid i\rangle\mid^2 \nonumber\\
&\times&
\delta(\omega-(E_{\nu}-E_i)).
\label{Eq:XPS}
\end{eqnarray}
Similarly, the  absorption spectrum $\kappa(\omega)$ is  given by
\begin{eqnarray}
\kappa_{\bf e_i, k_i}(\omega)&=& \frac{1}{Z} \sum_{i,\nu} e^{-\beta E_i} \mid \langle\nu\mid \hat D_{\bf k_i}({\bf e_i})\mid i \rangle\mid^2 \nonumber\\
&\times&\delta(\omega-(E_\nu-E_i)).
\label{Eq:XAS}
\end{eqnarray}

The RIXS cross-section $R(\omega_i,\omega_f)$ in the Kramers-Heisenberg representation is
\begin{eqnarray}
&&R({\bf e_i,e_f,k_i,k_f},\omega_i,\omega_f)= \frac{1}{Z} \sum_{i,f} e^{-\beta E_i} \nonumber\\
&\times&\left\vert \sum_\nu \frac{\langle f \mid \hat D_{\bf k_f}^*({\bf e_f})\mid\nu\rangle \langle\nu\mid \hat D_{\bf k_i}({\bf e_i})\mid i\rangle}{\omega_i-(E_\nu-E_i)-i \Gamma}\right\vert^2 
\nonumber\\
&\times&\delta(\Omega-(E_f-E_i)).
\label{Eq:RIXS}
\end{eqnarray}
This formula is the dominant resonant diagram to lowest order in the fine structure constant $\alpha = e^2/\hbar c$ \cite{Raman,RIXS}. In these expressions, $Z$ is the partition function, $E_{\{i,f\}}$, $\mid\{i,f\}\rangle$ are the initial and final state energies and eigenstates, respectively, $E_{\nu}, \mid \nu\rangle$ are the XPS/ARPES and XAS final states or RIXS intermediate core-hole states, $\Gamma$ is the core-hole decay rate, and ${\bf e_{\{i,f\}}}$, ${\bf k_{\{i,f\}}}$, and $\omega_{\{i,f\}}$ are the incident and scattered x-ray polarization vectors, momenta, and energies, respectively, with $\Omega=\omega_i-\omega_f$ is the energy loss. The dipole transition operator $D_{\bf k_i}({\bf e_i})$ for XPS/ARPES represents the removal of a valence electron, while for XAS/RIXS it represents the transition of a core electron up to the valence. Note that in these cases the cross-sections are written in terms of matrix elements squared, determined by incident and scattered polarization dipole transitions, and density of states plus energy and momentum conservation.

These applications are based on full matrix diagonalization of the generalized Hamiltonian
\begin{equation}
\hat H_{full}=\hat H_{int}+\hat H_{CEF} + \hat H_{SO}+\hat H_{KE}
\label{Eq:full}
\end{equation}
where the terms in the Hamiltonian written in second quantized operators $c_{\mu,i,\sigma}$ ($c^\dagger_{\mu,i,\sigma}$) that annihilate (create) an electron in orbital $\mu$ and unit cell $i$ with spin $\sigma$ are as follows: 

The Coulomb interaction
\begin{eqnarray}
\hat H_{int}&=&\frac{1}{2} \sum_i\sum_{\alpha,\beta,\gamma,\delta}\sum_{\sigma,\sigma'} U_{\alpha,\beta,\gamma,\delta}\nonumber\\ &\times&c^{\dagger}_{i,\alpha,\sigma} c^{\dagger}_{i,\beta,\sigma'} c_{i,\gamma,\sigma'} c_{i,\delta,\sigma}
\label{Eq:Hint}
\end{eqnarray}
represents generalized local two-body interactions that can include both core and valence electrons, while the sum of single particle terms can be grouped together as
\begin{equation}
\hat H_{CEF} + \hat H_{SO}+\hat H_{KE} =
\sum_{i,j}\sum_{\sigma,\sigma'}\sum_{\mu,\nu}
t_{i,j,\sigma,\sigma'}^{\mu,\nu} c^\dagger_{i,\mu,\sigma}c_{j,\nu,\sigma'}
\end{equation}
with the tensor $t$ having site energies, crystal fields and spin-orbit contributions (for $i=j$ terms), as well as hybridization among orbitals ($i\ne j$).

\begin{figure}
\includegraphics[width=\columnwidth]{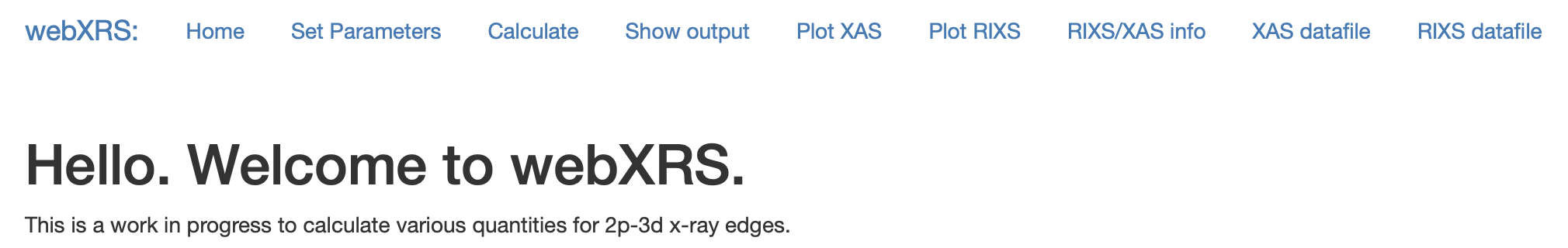}
\caption{\label{Fig1}
Home screen for the {\tt webXRS} application.}
\end{figure}

While in principle the orbital index $\mu$ can be chosen according to convenience as any complex set of basis states, in the applications {\tt webXRS} and {\tt webARPES} we choose atomic orbitals as basis states and take the terms of the Hamiltonian as Slater-Koster parameters, entered by the user. These parameters $F_0, F_2, \dots, G_1, G_3, \dots$ correspond to ascending multipole Coulomb matrix elements between various combinations of orbitals, and control the various density-like and Hund's-like interactions among valence and core electrons. These are explained in detail in various textbooks. In addition, the user may examine their effect in the plotted spectra but also in the output datafiles available for both viewing and downloading.

The dipole operator is evaluated in the Coulomb or ``transverse" gauge $\nabla\cdot {\bf A}=0$, where ${\bf A}$ is the light vector potential, meaning that the light polarization ${\bf e}$ lies in a plane perpendicular to the direction of propagation. For the application \url{webXRS} we utilize the dipole approximation to represent the dipole operator. Currently \url{webXRS} is configured such that
the incident photon creates a core-hole in the $2p$ shell and removes one from the valence $3d$ shell, specifically targeting $L$-edges of $3d$ transition metals. The light polarization vector ${\bf e}$ selects-out orbitally dependent transitions, viz.,
\begin{eqnarray}
\hat D_{2p,3d}({\bf e}) &= &\sum_{m,m',\sigma} d_{m,m'} ({\bf e})c^{\dagger}_{2p,m',\sigma} c_{3d,m,\sigma}\nonumber\\
d_{m,m'}({\bf e})& = &d ~~n_{m-m'}({\bf e}) ~~c^1(l=1,m';l=2,m)\nonumber\\
d &=& \int_0^{\infty} dr r^3 R_{3,2}(r) R_{2,1}(r)
\end{eqnarray}
where $c^1(l,m;l'm')$ is a Gaunt coefficient. Here $d$ can be neglected as it is a simple constant given by an integral of radial atomic wavefunctions and only sets an overall scale. 

\begin{figure}
\includegraphics[width=\columnwidth]{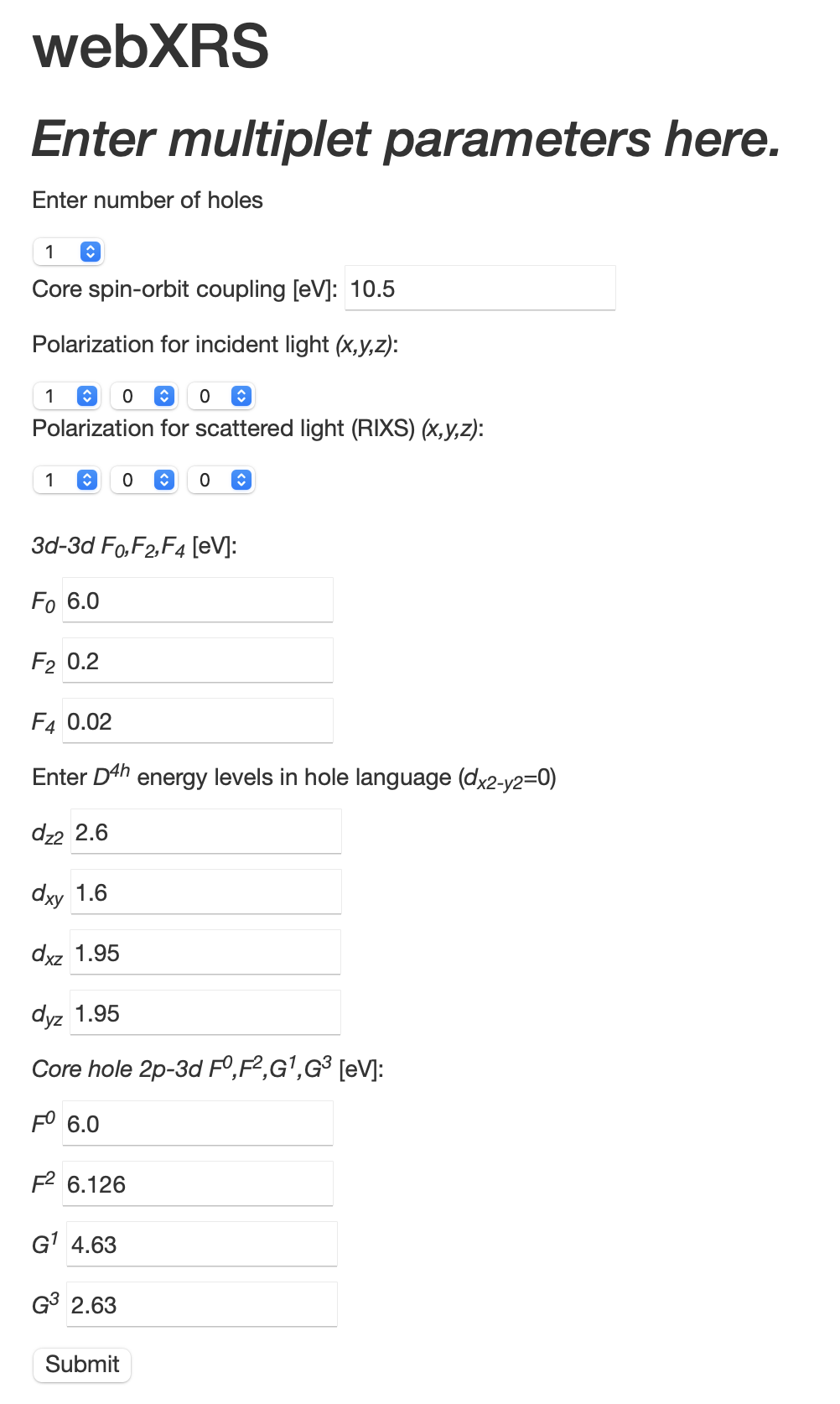}
\caption{\label{Fig2}
How parameters are set.}
\end{figure}
For the application {\tt webARPES}, the final state for photoemission is taken to be a plane wave having energy $\omega$ and wavevector ${\bf k}$ arriving at a detector. This gives

\begin{figure}
\includegraphics[width=\columnwidth]{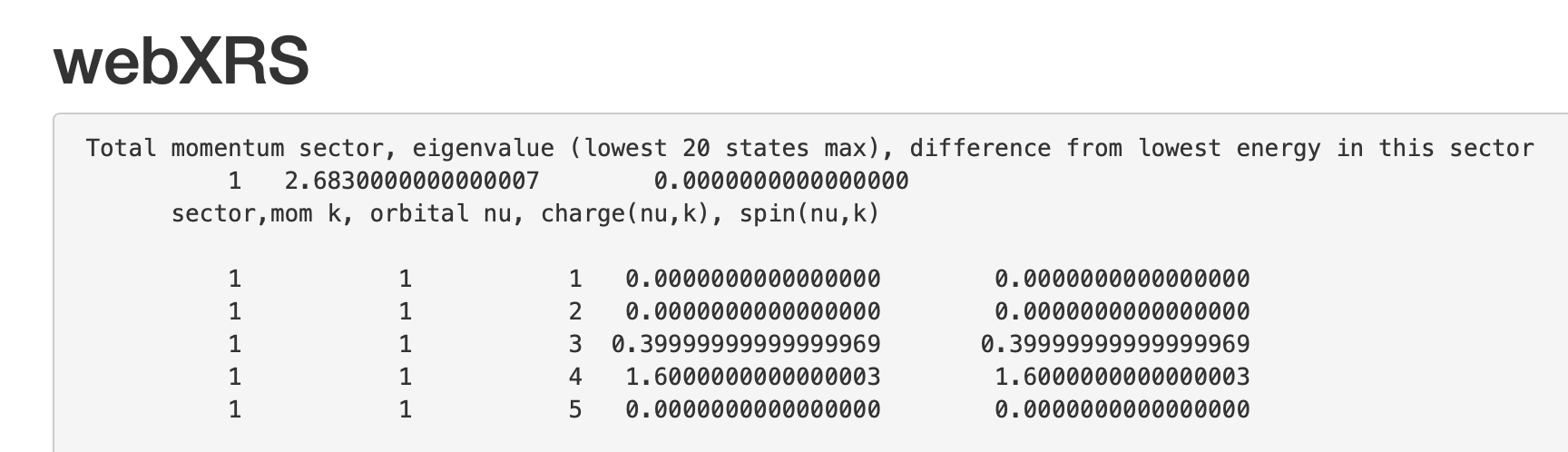}
\caption{\label{Fig3}
Eigenstate information of the $d^2$ hole system upon code execution.}
\end{figure}

\begin{eqnarray}
\hat D_{{\bf k},\nu}({\bf e})&=& \sum_{\sigma,l,m} c_{k,\sigma}^\dagger c_{\nu,\sigma} Y_{l,m}(\theta_k,\phi_k) d_{l,m;l_{\nu},m_{\nu}}({\bf e},k)\nonumber\\
d_{l,m;l_{\nu},m_{\nu}}({\bf e},k) &=& c^1(l,m; l_{\nu},m_{\nu}) d_{l;n_{\nu},l_{\nu}}(k) n_{m-m_{\nu}}({\bf e})\nonumber\\
d_{l;n_{\nu},l_{\nu}}(k) &=& 4\pi(-i)^l\int dr r^3 j_l(kr) R_{n_{\nu},l_{\nu}}(r),
\end{eqnarray}
with spherical harmonic $Y_{l,m}$ and $j_l$ spherical Bessel functions. Here, we assume that the quantum index $\nu$ denotes a valence state having the same momentum and spin as the outgoing plane wave but associated with orbital index $\nu$. This is not just a simple Fourier transform of matrix elements since the orbital wavefunction overlap with the expansion of Bessel functions affects the overall photoemission matrix elements.

\begin{figure}
\includegraphics[width=\columnwidth]{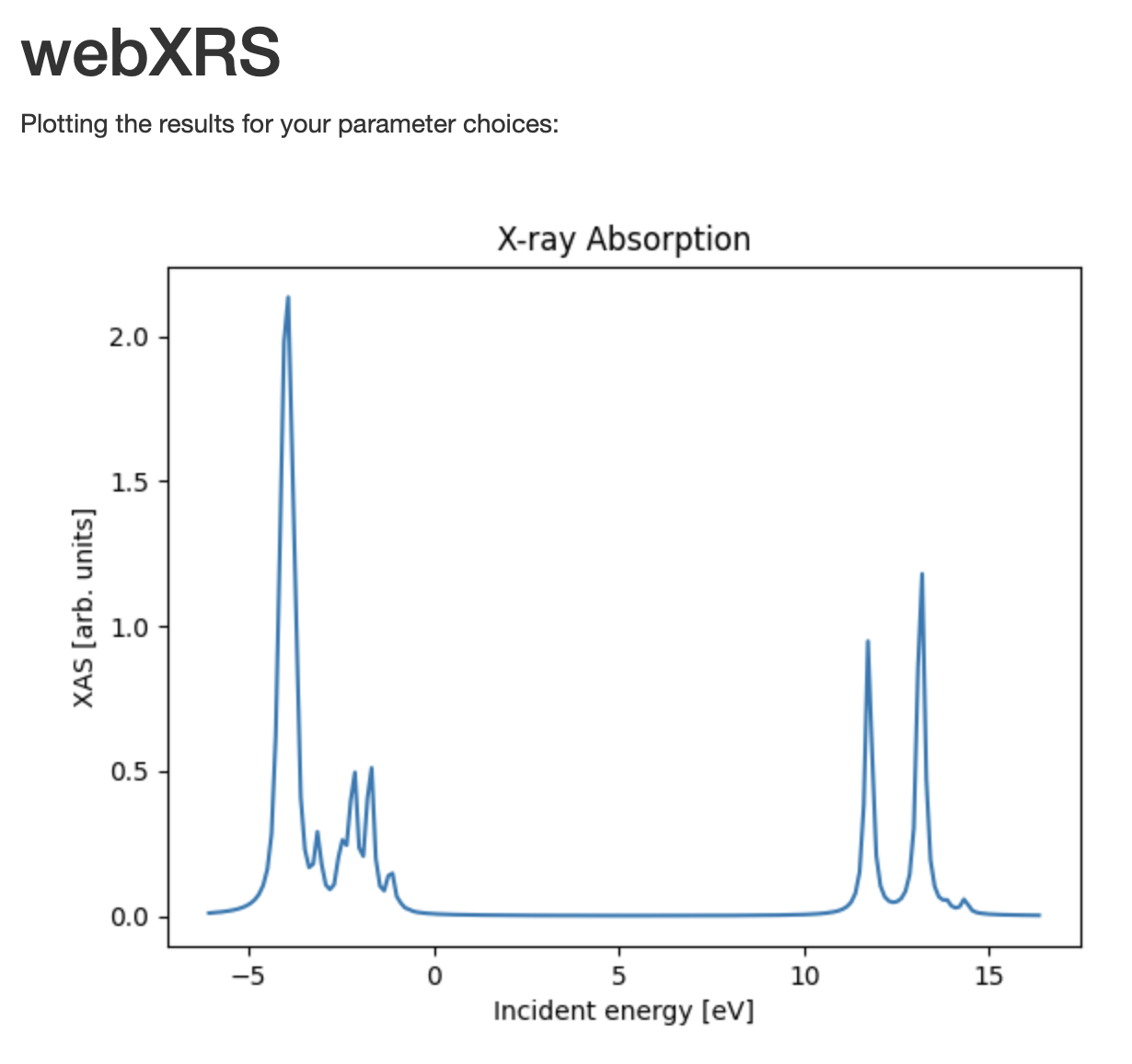}
\caption{\label{Fig4}
Displayed XAS for $d^2$ system obtained from the application.}
\end{figure}

\section{Applications}

Both \url{webXRS} and \url{webARPES} diagonalize the full Hamiltonian in Eq. (\ref{Eq:full}) to obtain eigenvalues and eigenvectors for the states appearing in Eqs. (\ref{Eq:XPS}-\ref{Eq:RIXS}). The diagonalization and assembly routines are performed on the server so no other configuration is needed apart from setting the input parameters, obviating the need for code download and installation. This of course limits what each application may perform, but gives the user a simple, interactive way to learn about spectroscopy.

\begin{figure}
\includegraphics[width=\columnwidth]{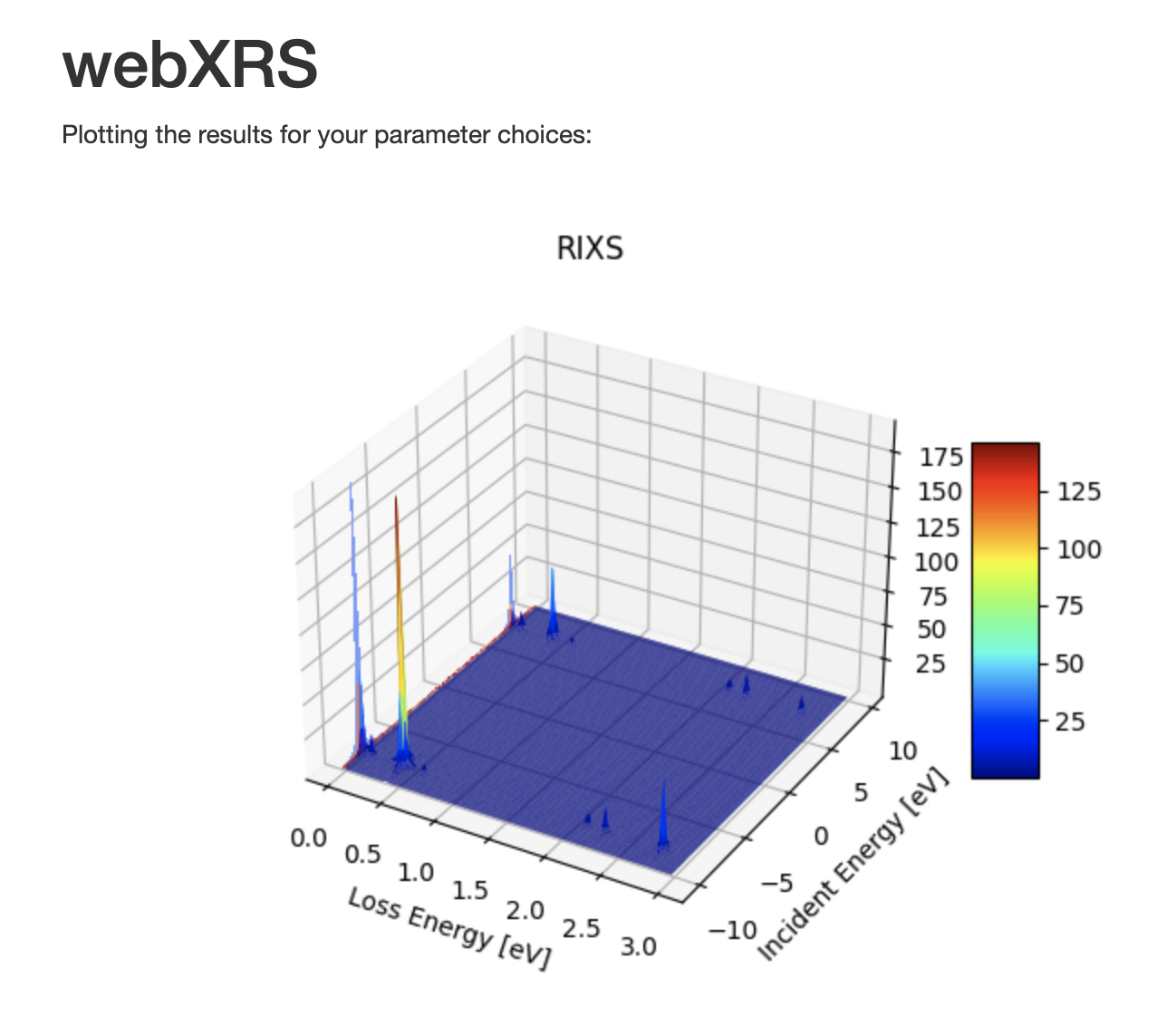}
\caption{\label{Fig5}
Displayed RIXS for $d^2$ system obtained from the application.}
\end{figure}

\subsection{\tt webXRS}
Currently, {\tt webXRS} (see Fig. \ref{Fig1}) is configured to calculate transition metal $L$-edge XAS and RIXS for a single unit cell composed on a single $3d^n$ transition metal having Coulomb parameters $F_0,F_2,$ and $F_4$ setting the valence interaction matrix elements, and separate Slater values for the $2p-3d$ core-valence interactions $F^0,F^2,G^1$ and $G^3$. The Hilbert space dimension for the initial states is given by the binomial coefficient $C_{10}^{n}$, with $n$ representing the number of holes, and $6 \times C_{10}^{n-1}$ for the XAS final states or RIXS intermediate states. The user may navigate the application by choosing the action tabs on the top of the entry pages. The home page gives information of code capability updates as well as a list of features that are planned in future versions of the application. 

Through the {\it Set Parameters} tab, crystal field levels (tesseral harmonic energies) and core spin-orbit coupling are input, as are the incoming and outgoing light polarization orientations (see Fig. \ref{Fig2}). An input file is then created for code execution, residing on the host server.

In principle, all of these parameters can be determined from various first-principle (i.e., density functional theory) approaches. However, in this application they can be simply taken as fit parameters. Consideration of the parameters entered follows their effect on the calculated eigenspectra, as evidenced in the so-called Tanabe-Sugano diagrams for each $d^n$ system for the valence electrons: the Slater parameters governed the splitting of the "multiplets", while the crystal field levels change the energy position of the orbitals as well as eigenstate degeneracies. The core spin-orbit interaction can be calculation via atomic orbitals, but again here it is chosen to simply give the splitting between the $2p_{1/2}$ and $2p_{3/2}$ core levels and the corresponding $L_2$ - $L_3$ edge separation. Finally, the core-valence interactions "shuffle" spectral weight within each specific edge due the orbital nature of these interactions. The reader (and user) is directed to Ref. \onlinecite{KotanideGroot} for details.

Upon execution of the application by utilizing the {\it Calculate} tab, the user may view the energy and orbital and spin occupations for the lowest energy levels, the values of the intra-, inter-, and Hund's interactions $U, U', J$, respectively, and view plots of XAS and RIXS maps as a function of incoming photon energy. Data files may be downloaded by the user for further processing.

The eigenstate information of the $d^n$ system is shown via the {\it Show Output} tab in Fig. \ref{Fig3} whereby the user may inspect information of the eigenenergies and eigenstates of the lowest twenty states. The orbital spin and charge decompositions are given for each eigenstate.

\begin{figure}
\includegraphics[width=\columnwidth]{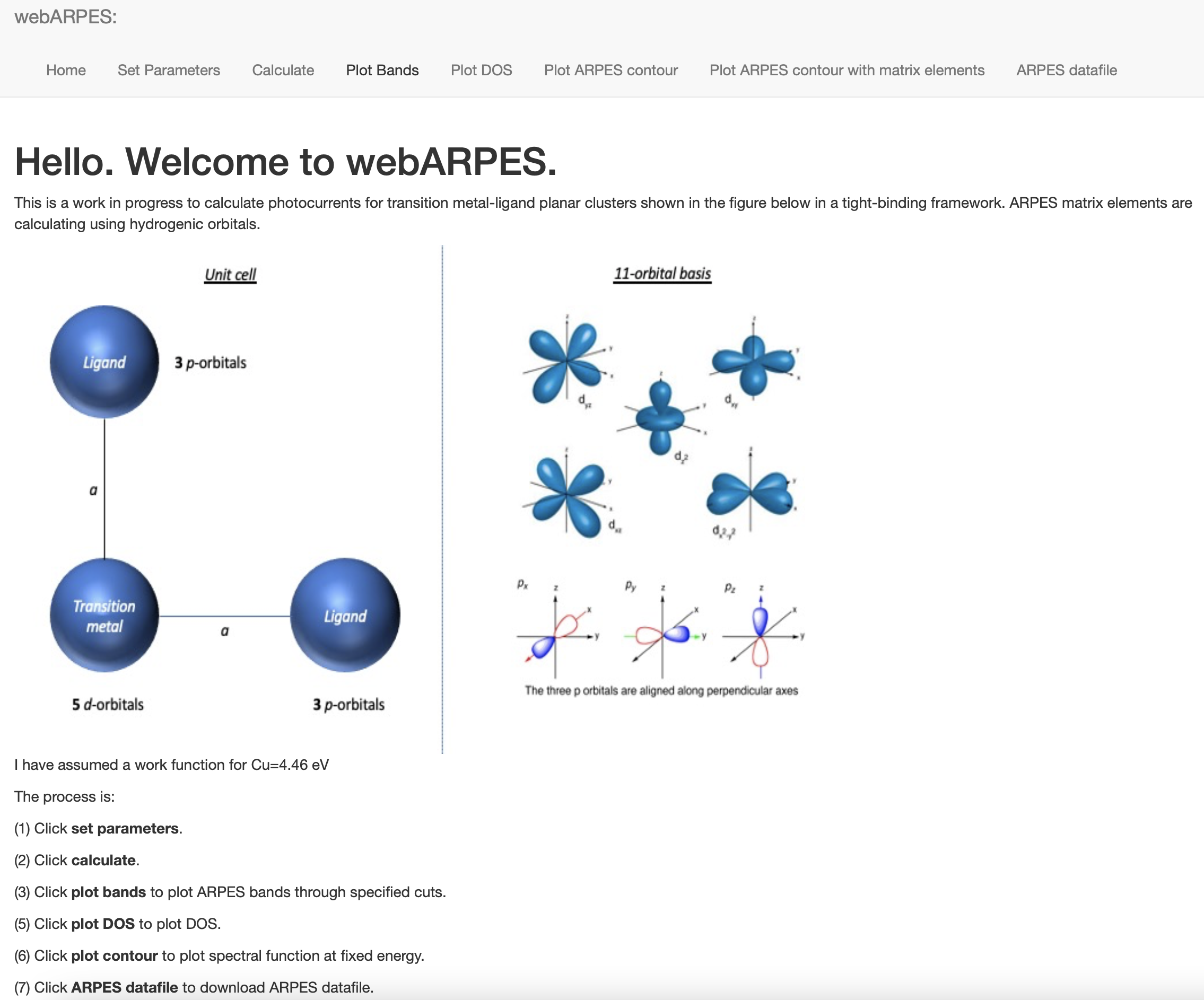}
\caption{\label{Fig6}
Homepage for the {\tt webARPES} application.}
\end{figure}

The XAS spectra can then be plotted using the {\it Plot XAS} tab in the application. The corresponding tab delimited two column datafile can be downloaded via the menu bar {\it XAS Datafile}.

The RIXS spectrum can then be plotted as a RIXS map (function of incident and scattered photon energies) using the {\it Plot RIXS} tab in the application. A resulting 3D plot (see Fig. \ref{Fig5}) highlights the spectral intensities of both the $L_{2,3}$ edges that can be correlated with the XAS absorption plots (Fig. \ref{Fig4}) for the same incoming photon polarization. A corresponding tab delimited three column file can be downloaded via the menu bar {\it RIXS Datafile}. With the downloaded datafiles the user may choose to examine the results using their own local plotting software tools.

Thus, \url{webXRS} can be a useful pedagogical tool whereby users can learn about multiplets and x-ray absorption and resonant inelastic x-ray scattering formalism using simple exact diagonalization codes that do not require downloading or user installation. The code execution is fast enough to allow for rapid access that may be useful as input to more complex computational frameworks.

\begin{figure}[b]
\includegraphics[width=\columnwidth]{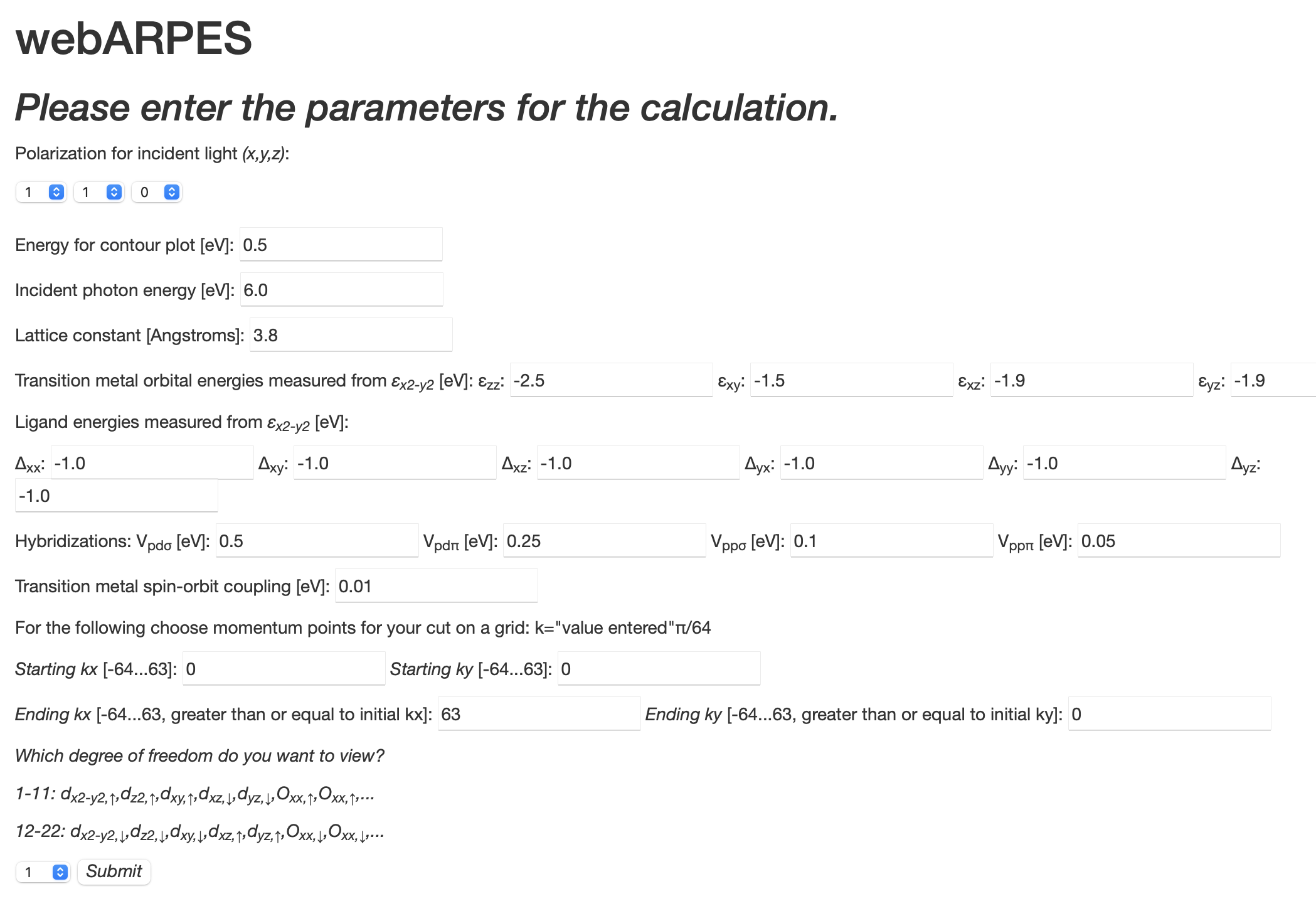}
\caption{\label{Fig7}
Data entry page using the {\it Set Parameters} tab for tight-binding parameters as well as photon configuration.}
\end{figure}

\subsection{{\tt webARPES}}
The ARPES application {\tt webARPES} is currently configured for a fixed number of unit cells $(128 \times 128)$ composed of a single $d-$transition metal ion in a square planar lattice of $p-$ligand orbitals, giving an 11-orbital unit cell. The home page entry point similar to the {\tt webXRS} application is shown in Fig. \ref{Fig6}, indicating the orbital basis used for the calculation as well as general application information. An updated posting of planned features and new activities is shown along with the general purpose of the application.

Upon execution of the application (again hosted on the server), the user may view the energy eigenvalues (bands) plotted as a function of momentum along different Brillouin zone (BZ) directions, plot the full and orbitally-resolved density of states (DOS), fixed energy contours throughout the BZ, and compare the effect of photon dipole transition matrix elements. The data files may be downloaded for further processing.

Tight-binding parameters and spin-orbit coupling are entered by the user, and the incident light energy, momentum (in units of lattice constant $a$), and polarization orientation are to be specified as well (see Fig. \ref{Fig7}). In the current version of the application, no Coulomb matrix elements are included so that a basis of single-particle Slater determinants can be used. Therefore the calculation diagonalizes a $22 \times 22$ matrix for each momentum point on the grid, given by blocks of $11 \times 11$ matrices for each spin orientation
\begin{widetext}
\begin{align}
\begin{pmatrix}
\epsilon_{x^2-y^2} & 0 & 0 & 0 & 0 & -t_{pd}s_x & 0 & 0 & 0 & t_{pd}s_y & 0 \\
0 & \epsilon_{z^2} & 0 & 0 & 0 & t_{pdz}s_x & 0 & 0 & 0 & t_{pdz}s_y & 0 \\
0 & 0 & \epsilon_{xy} & 0 & 0 & 0 & t_{pdxy}s_x & 0 & t_{pdxy}s_y & 0 & 0 \\
0 & 0 & 0 & \epsilon_{xz} & 0 & 0 & 0 & t_{pdxz}s_x & 0 & 0 & 0 \\
0 & 0 & 0 & 0 & \epsilon_{yz} & 0 & 0 & 0 & 0 & 0 & t_{pdyz}s_y  \\
-t_{pd}s^*_x & t_{pdz}s^*_x & 0 & 0 & 0 & \Delta_{xx} & 0 & 0 & -t_{pp\pi}c_xc_y & t_{pp\sigma}s_xs_y & 0 \\
0 & 0 & t_{pdxy}s^*_x & 0 & 0 & 0 & \Delta_{xy} & 0 & t_{pp\sigma}s_xs_y & -t_{pp\pi}c_xc_y & 0 \\
0 & 0 & 0 & t_{pdxz}s^*_x & 0 & 0 & 0 & \Delta_{xz} & 0 & 0 & t_{ppz\pi}c_xc_y \\
0 & 0 & t_{pdxy}s^*_y  & 0 & 0 & -t_{pp\pi}c_xc_y & t_{pp\sigma}s^*_xs^*_y & 0 & \Delta_{yx} & 0 & 0 \\
t_{pd}s^*_y & t_{pdz}s^*_y & 0 & 0 & 0 & t_{pp\sigma}s^*_xs^*_y & -t_{pp\pi}c_xc_y & 0 & 0 & \Delta_{yy} & 0 \\
0 & 0 & 0 & 0 & t_{pdyz}s^*_y & 0 & 0 & t_{ppz\pi}c_xc_y & 0 & 0 & \Delta_{yz} \nonumber
\end{pmatrix}
\end{align}
for a $\{\epsilon_{x^2-y^2}, \epsilon_{z^2}, \epsilon_{xy}, \epsilon_{xz}, \epsilon_{yz}, O_{xx}, O_{xy}, O_{xz}, O_{yx}, O_{yy}, O_{yz}\}$ basis.
Here $s_{x,y}=-2i\sin(k_{x,y}a/2)$, $c_{x,y}=2\cos(k_{x,y}a/2)$, and $\epsilon$'s and $\Delta$'s denote the transition metal site and charge transfer energies, respectively. These degenerate blocks are coupled when spin-orbit coupling is initiated in the calculation.
\end{widetext}

\begin{figure}
\includegraphics[width=\columnwidth]{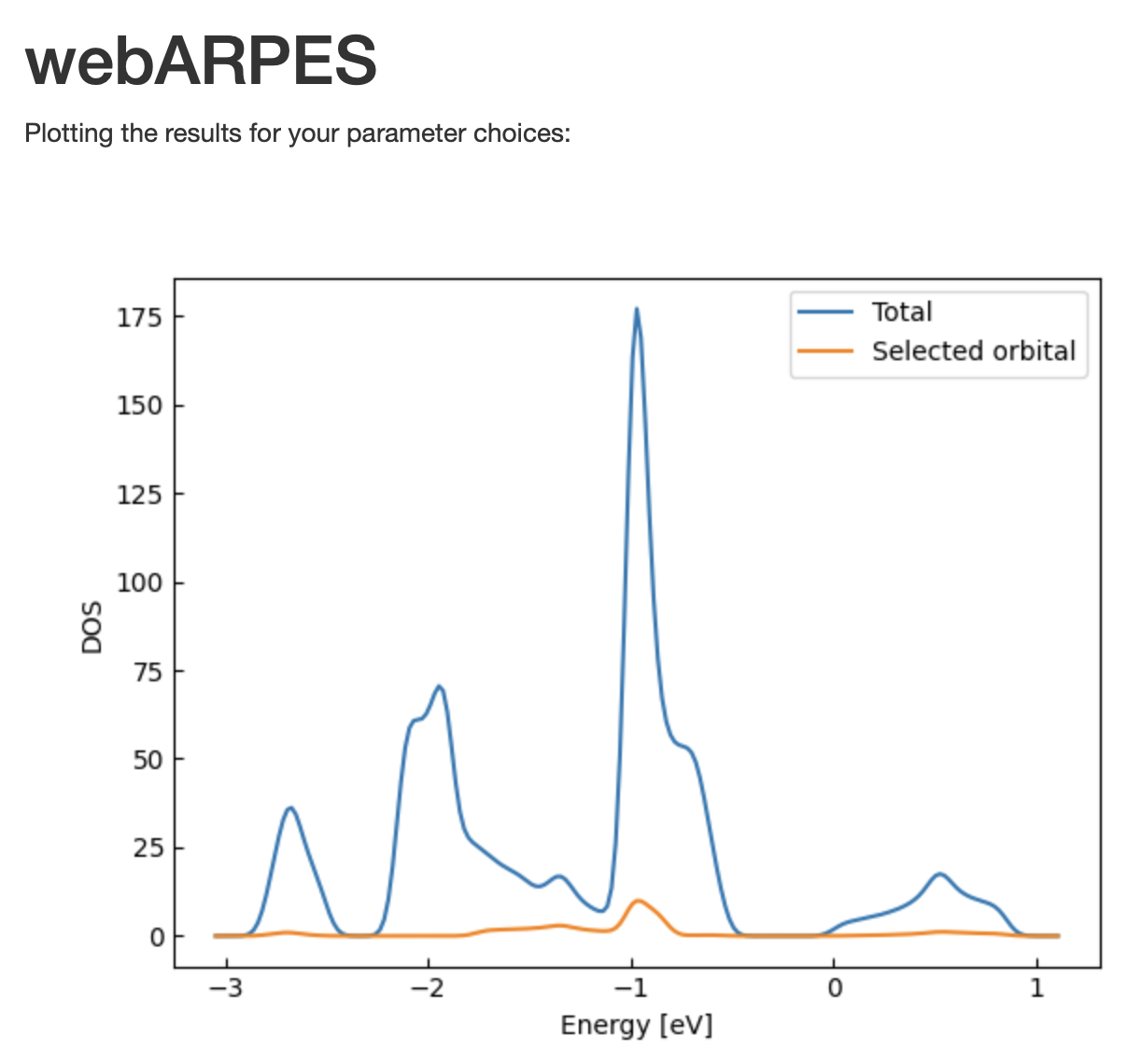}
\caption{\label{Fig8}
Total and orbitally projected density of states.}
\end{figure}

On the same page a user makes a selection of momentum points of the resulting band structure (eigenvalues) for plotting purposes, as well as which orbitals can be projected (components of the eigenvectors) for examination in both the bandstructure and the density of states. These are plotted in the {\it Plot DOS} tab of the application (Fig. \ref{Fig8}).

\begin{figure} 
\includegraphics[width=\columnwidth]{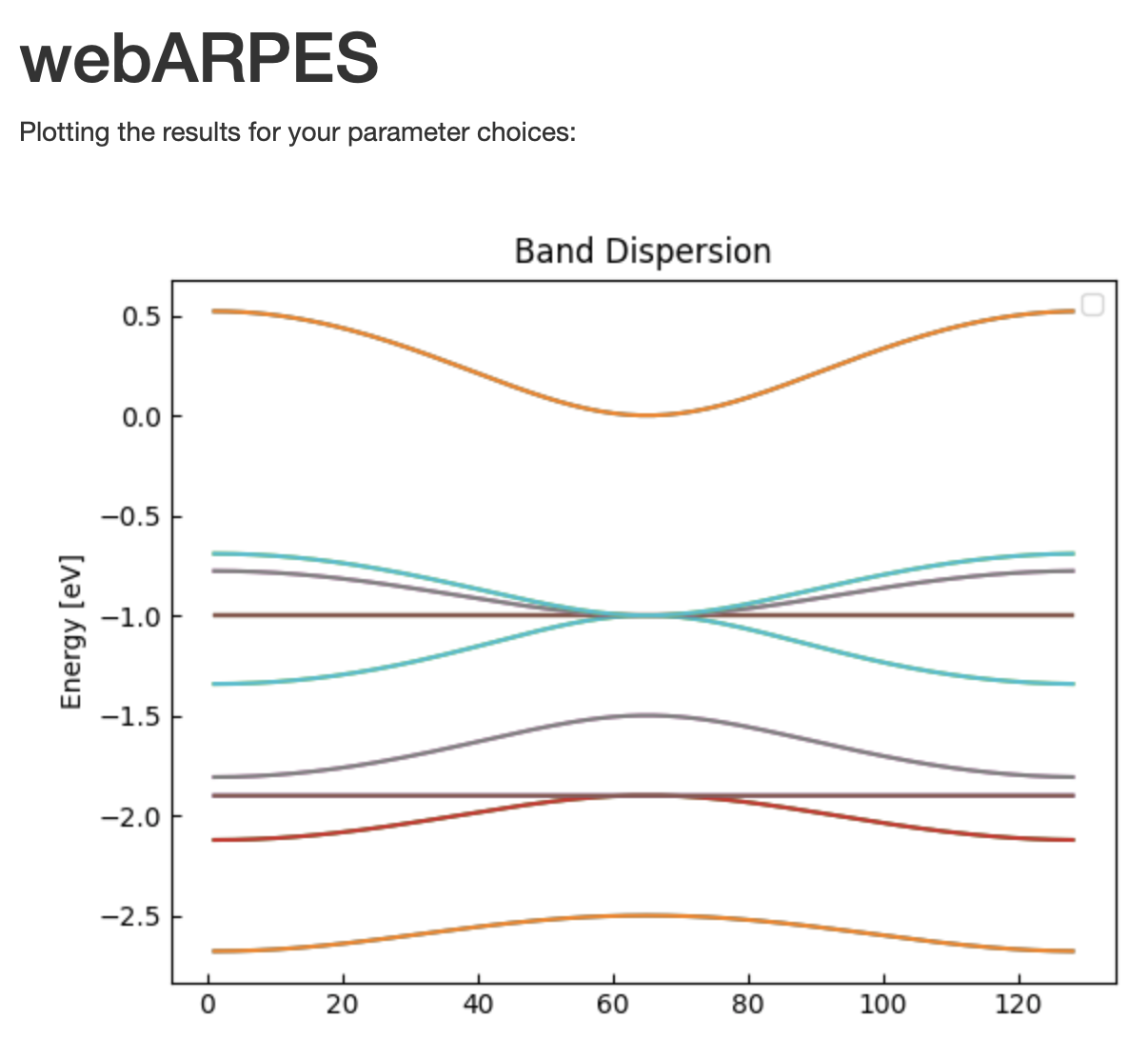}
\caption{\label{Fig9}
Energy eigenvalues plotted along the specified momentum cut directions in the Brillouin Zone. The $x$-axis corresponds to the points requested by the user along the cut in units of $\pi/a$, with $a$ the lattice constant.}
\end{figure}

The {\it Plot Bands} tab displays the sorted eigenvalues for the momentum cuts specified by the user in the {\it Set Parameters} tab (Fig. \ref{Fig9}). The user may return to the {\it Set Parameters} tab to choose other cut directions or orbital projections to plot.

Finally, the projection of the bands with and without the inclusion of dipole matrix elements can be plotted within the {\it Plot ARPES Contour} and {\it Plot ARPES Contour with matrix elements} tabs, respectively, using the parameters specified by the user in the {\it Set Parameters} tab. Both plots show the bands across a repeated zone scheme to highlight the effects of the matrix elements due to the multiorbital nature of the configuration.

\begin{figure}[b] 
\includegraphics[width=\columnwidth]{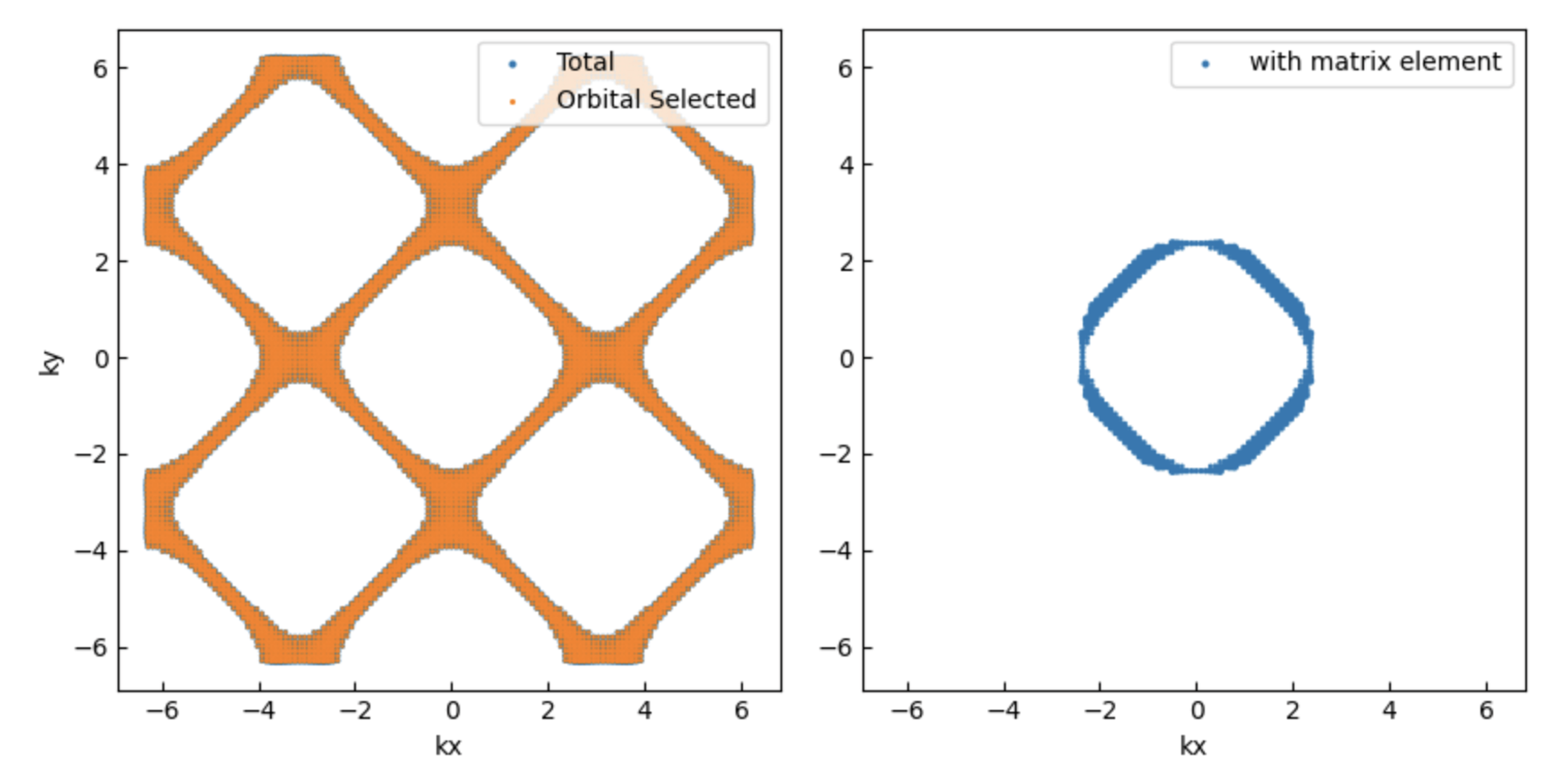}
\caption{\label{Fig10}
ARPES energy contours without (left) and with (right) the inclusion of matrix elements.}
\end{figure}

\section{Summary}
In summary, a web-based application is introduced that calculates a number of photon-based spectroscopies using different approaches for quantum materials. While the calculations are simple and straightforward but limited in scope, it is envisioned that this tool can help new students and users in the field of spectroscopy to become familiarized with simple concepts without the associated difficulties of application installation or configuration. The software is easily extendable to other systems, such as to larger clusters and unit cells, with the requisite cost in local server memory and access.

\acknowledgements
This work was supported by the U.S. Department of Energy (DOE), Office of Basic Energy Sciences,
Division of Materials Sciences and Engineering, specifically, through the Theory Institute for Materials and Energy Sciences (TIMES) program.

\newcommand{\bibtitle}[1]{\textit{#1},}
\newcommand{\bibvol}[1]{\textbf{#1}}

\end{document}